\documentclass[12pt,onecolumn,twoside]{osajnl}

\usepackage{soul}
\DeclareRobustCommand{\rev}[1]{{\sethlcolor{white}\hl{#1}}} 

\journal{ao} 

\setboolean{shortarticle}{true} 

\usepackage{gensymb}
\usepackage{amsmath,amssymb}
\usepackage{textcomp}
\graphicspath{{./Figures/}{./}}
\DeclareGraphicsExtensions{.pdf,.jpeg,.png,.eps}

\title{First industrial-grade coherent fiber link for optical frequency standard dissemination}

\author[1]{F. Guillou-Camargo}
\author[1]{V. M{\'e}noret}
\author[2,3]{E. Cantin}
\author[2]{O. Lopez}
\author[2,4]{N. Quintin}
\author[4]{E. Camisard}
\author[5]{V. Salmon}
\author[5]{J.-M. Le Merdy}
\author[3,6]{G. Santarelli}
\author[2]{A. Amy-Klein}
\author[3]{P.-E. Pottie}
\author[1]{B. Desruelle}
\author[2]{C. Chardonnet}

\affil[1]{Muquans, Institut d'Optique d'Aquitaine,Rue Fran{\c c}ois Mitterrand, 33400, Talence, France}
\affil[2]{Laboratoire de Physique des Lasers, Universit\'{e} Paris 13, CNRS, Villetaneuse, France}
\affil[3]{Laboratoire National de M\'{e}trologie et d'Essais-Systeme de R\'{e}f\'{e}rences Temps-Espace, Observatoire de Paris, Universit\'{e} PSL, CNRS, Paris, France }
\affil[4]{RENATER, Paris, France, 23-25 rue Daviel, 75013 Paris, France} 
\affil[5]{Syrlinks, 28 rue Robert Keller, ZAC des Champs Blancs, 35510 Cesson-S\'{e}vign\'{e}, France} 
\affil[4]{LP2N, Laboratoire de Photonique Num\'{e}rique et Nanosciences, Institut d’Optique Graduate School, rue Fran{\c c}ois Mitterrand, 33400 Talence, France}

\affil[*]{Corresponding author: bruno.desruelle@muquans.com}

 \dates{Compiled \today}

\ociscodes{(120.3930) Metrological instrumentation; (060.2360) Fiber optics links and 
subsystems; (140.0140) Lasers and laser optics; (120.5050) Phase measurement. }

\begin{abstract}
We report on a fully bi-directional 680~km fiber link connecting two cities for which the equipment, the set up and the characterization are managed for the first time by an industrial consortium. The link uses an active telecommunication fiber network with parallel data traffic and is equipped with three repeater laser stations and four remote double bi-directional Erbium-doped fiber amplifiers. We report a short term stability at 1~s integration time of $5.4\times 10^{-16}$ in 0.5~Hz bandwidth and a long term stability of $1.7\times10^{-20}$ at 65\,000 s of integration time. The accuracy of the frequency transfer is evaluated as $3\times 10^{-20}$. No shift is observed within the statistical uncertainty. We show a continuous operation over 5 days with an uptime of 99.93$\%$. This performance is comparable with the state of the art coherent links established between National Metrology Institutes in Europe. It is a first step in the construction of an optical fiber network for metrology in France, which will give access to an ultra-high performance frequency standard to a wide community of scientific users. 
\end{abstract}
\setboolean{displaycopyright}{true}
\DeclareUnicodeCharacter{2212}{-}

\begin{document}

\maketitle

\section{Introduction}
Optical fiber links are an emerging technology for high resolution time and frequency transfer and comparison. Single mode optical fibers are a very promising medium to carry metrological signals because the guided propagation ensures excellent stability of the propagation delay, an unambiguous propagation path and a much higher signal and lower noise than propagation in free space over long hauls. 

In order to reach the lowest transfer instability, propagation noise has to be cancelled\,\cite{primas_fiber_1988}. Due to the excellent reciprocity of the phase accumulated forth and back, the noise can be measured using a round-trip signal\,\cite{ma_delivering_1994}. This can be done by retro-reflecting part of the signal, or by using a laser which is offset phase locked to the incoming signal\,\cite{foreman_coherent_2007}. This latter technique has numerous advantages, \rev{as the light injected in the backward direction is constant in power}. It is a key technique for the regeneration of laser light in a cascaded link, where the link length is sub-divided in smaller spans, allowing for a better noise rejection\,\cite{musha_coherent_2008}. In addition it provides the opportunity to deliver a useful signal to an end-user\,\cite{chiodo_cascaded_2015}. This technique is developed since 10~years in the optical and microwave domain\,\cite{lopez_cascaded_2010,fujieda_coherent_2010, fujieda_all-optical_2011} and was recently implemented on a long-haul link of $2\times710$~km, with 4 cascaded spans\,\cite{chiodo_cascaded_2015}. Cascaded link technique was used in Europe for clock comparisons between National Metrology Institutes (NPL, PTB and LNE-SYRTE)\,\cite{lisdat_clock_2016,guena_first_2017,delva_test_2017}.%

It motivates a wide community of scientists, from fundamental to applied physics and industry, to consider this technique for various applications\,\cite{riehle_optical_2017,takano_geopotential_2016,clivati_measuring_2016,matveev_precision_2013,argence_quantum_2015,clivati_vlbi_2017, marra_seismology_2017,baldwin_dissemination_2016}. \rev{Projects aiming at creating metrological networks at national and continental scale over telecommunication fiber networks are pushing the technology forward, as their implementation requires a high level of maturity}\,\cite{levi_lift-italian_2013, buczek_optime_2016,krehlik_clonets_2017,chardonnet_refimeve_2018}. \rev{Such metrological networks require the know-how to be transferred from academia to industrial partners who can in turn provide the scientific community with a high quality of service. As a first step towards such wide scale metrological network, we report here on the first long-haul coherent fiber link built, set-up, and managed by an industrial consortium, with supervision abilities integrated into the network operating center of a national research and education network.}%

The manuscript is organized as follows. First we discuss the scientific objectives and subsequent requirements of a sustainable metrological network, the way the fiber is accessed and the equipment to be built. Then we show one of the pieces of equipment we have developed for this network, an industrial-grade version of the repeater laser station\,\cite{chiodo_cascaded_2015}. We focus our discussion on the remote abilities and the supervision that was implemented to guarantee a proper operability by a network operation center. We report on the performance obtained using a dedicated test bench, and describe the implementation and operation of one link of the network, connecting two cities in France over 680~km of coherent fiber link. We show our experimental results and compare with the state-of-the-art.%

\section{The REFIMEVE+ project}
\begin{figure}[t]
 \centering
 \fbox{\includegraphics[width=.95\linewidth]{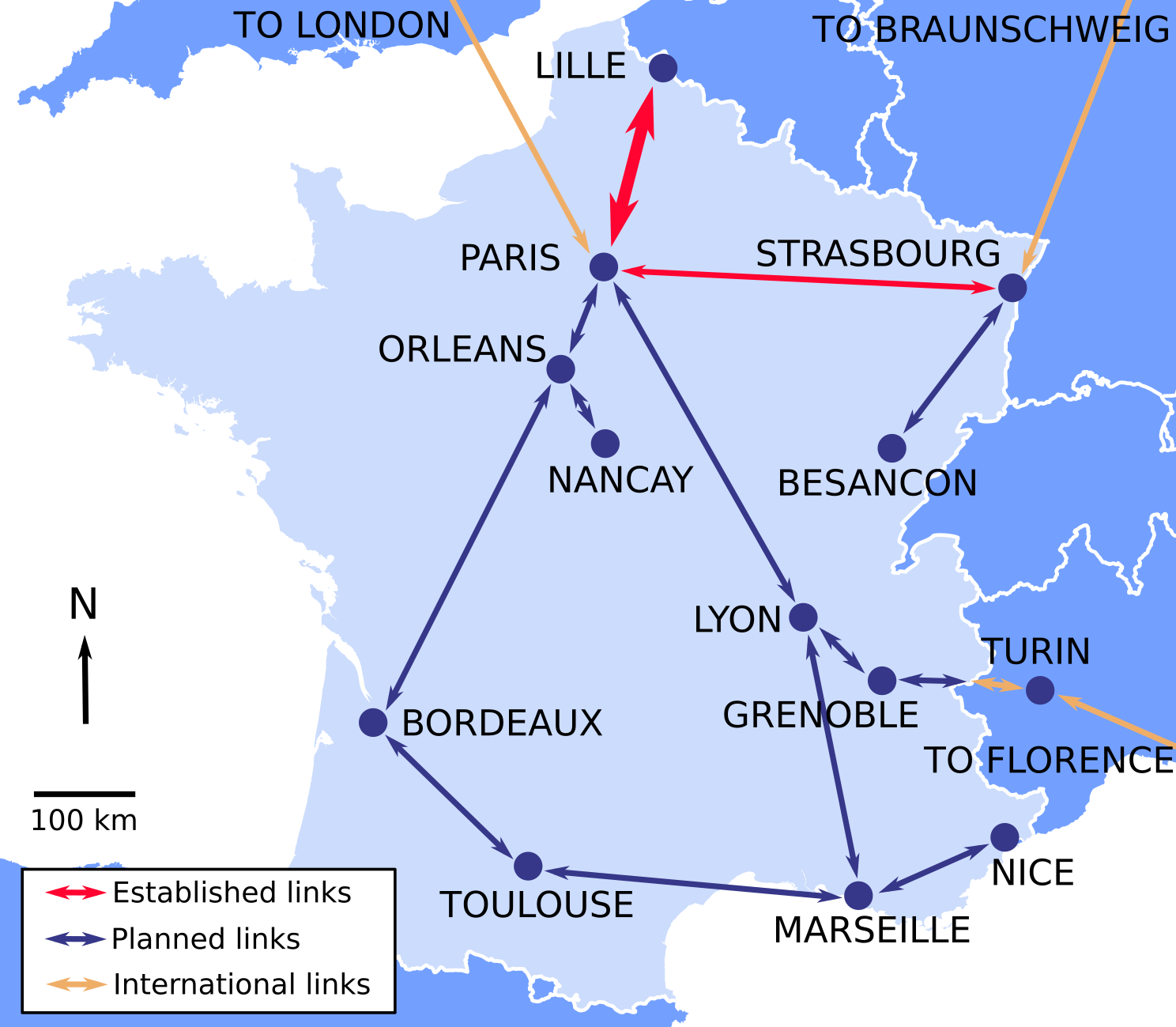}}
 \caption{Map of France with the REFIMEVE+ network. Red arrows: operational links. Blue arrows: planned links. Orange arrows: international cross-border links towards Braunschweig (Germany), London (UK) and Torino (Italy). In this paper, we focus on the link between Paris and Lille (thick red line). \label{fig:refimeve_carte}}%
 \end{figure}
 
The frame of the work presented in this manuscript is the REFIMEVE+ project\,\cite{chardonnet_refimeve_2018}. It aims at disseminating an ultra-stable optical frequency standard emitting at a wavelength in the C-band of the ITU grid over 4\,000~km of an active and operational telecommunication fiber network (see Figure\,\ref{fig:refimeve_carte}). The network is operated by RENATER, the French National Research and Education Network (NREN). The ultra-stable optical signal is generated at the French National Metrology Institute for Time and Frequency LNE-SYRTE. \rev{It consists in continuous-wave coherent light from a laser that is phase-locked to an ultra-stable cavity, and that} can be related to the absolute frequency of frequency standards using optical frequency combs. The ultra-stable signal has to be disseminated to about 20 research laboratories in France, for applications ranging from comparison of optical clocks and frequency standards to tests of advanced physics theories such as temporal variation of fundamental constants and general relativity, ultra-high resolution atomic and molecular spectroscopy, radio-astronomy and applied physics topics like the detection of atmospheric polluants. At the user end, the ultra-stable signal can be used directly, or with the help of an optical frequency comb.%

The scientific requirements are a short-term relative frequency stability of the optical signal at the level of $10^{-15}$ at 1~s averaging time, and below $10^{-19}$ at one day. The accuracy of the transferred signal must be limited by the frequency standards themselves. Finally, the up-time must be as close as possible to 100$\%$, in order to provide an efficient service to the scientific community.%

The fiber access is provided by RENATER. The metrological signal propagates in the same fiber as telecom data traffic using the Dense Wavelength Division Multiplexing (DWDM) technology\,\cite{kefelian_high-resolution_2009}. REFIMEVE+ uses the channel \#44 centered at 194.4~THz (1542.14~nm). This approach is essential for two reasons. First it reduces the costs as the fiber is shared and as many maintenance operations are handle by an operating center. \rev{Second it enables the operation of a scientific instrument on the long term, with an uptime as close as possible to 100\%.}%

Thermal and acoustic fluctuations in the fiber introduce phase noise which deteriorates the frequency stability of the metrological signal. To overcome this limitation, an active noise compensation is implemented, the round-trip phase fluctuations in the fiber being measured and corrected in real time. This requires the metrological signal to be bi-directional for optimal noise rejection. Any telecom or network equipment is by-passed thanks to bi-directional Optical Add/Drop Multiplexers (OADM). Channel \#44 is added and dropped in the internet network with 0.2~dB insertion loss for the data traffic. The OADM's insertion loss for the metrological signal is about 0.8~dB.%

Moreover the signal attenuation is about $0.3$~dB/km in average in-field, because of losses both in the fiber and on multiple optical interfaces. In every telecom shelter where a uni-directional EDFA has to be bypassed, a bi-directional EDFA amplifies the metrological signal. However, the gain of these amplifiers has to be kept quite low (typically 12-21~dB) to prevent self-oscillations due to parasitic Fabry-P{\'e}rot cavity effects. Therefore losses cannot be fully compensated with the bi-directional EDFAs. For long-range fiber links, better amplification techniques such as Fiber Brillouin amplification or regeneration techniques, for example using the repeater laser stations (RLS) described below, are used\,\cite{raupach_brillouin_2015,chiodo_cascaded_2015}. For the REFIMEVE+ project, the solution is to setup RLS along the link, every 300-500~km depending on the link noise and the link loss net budget. This solution allows us to set up cascaded links with better noise rejection and to build a fiber network which is both robust and reliable.%

Beyond the repeater laser stations and the bi-directional EDFAs, building a nation-wide network of 4\,000 km requires other specific modules, such as multi-user stations, central hubs, and end-user modules. For that purpose a specific technological development was conducted by Laboratoire de Physique des Lasers and co-workers. Technology transfer has been implemented to develop industrial grade equipment, a necessity for a robust and reliable operation of the REFIMEVE+ network. Not only manufacturing, but the whole process of installing, powering up, optimizing and operating the link in-field must be carried out according to industry standards so that REFIMEVE+ becomes a large effective and productive research infrastructure, with upscale capacity.%

Two key devices have been developed and used in the work presented here: the bi-directional amplifiers and the repeater laser stations. The bi-directional amplifiers have been developed by Keopsys\,\cite{keopsys_bidir_2018}. \rev{The minimal input power is -70~dBm and they achieve a noise figure below 5~dB with input signals as low as -50~dBm. Even with strong input signals, output power is limited by saturation to +5~dBm to comply with the operational requirements of the network.} This equipment has been conceived to incorporate two bi-directional amplifiers in one 19'' rack of 1~U height and 250~mm depth. This is compliant with the specifications of the remote shelters and nodes on the network. Moreover they are remotely controlled through IP access and included into the supervision and monitoring software discussed below (subsection\,\ref{SubSec:Supervision}). The RLSs have been developed by the French company Muquans with a first step of network supervision, and a detailed presentation is shown below.%

\section{The industrial-grade repeater laser station}
\subsection{Concepts}

\begin{figure}[t]
 \centering
  \fbox{\includegraphics[width=0.95\linewidth]{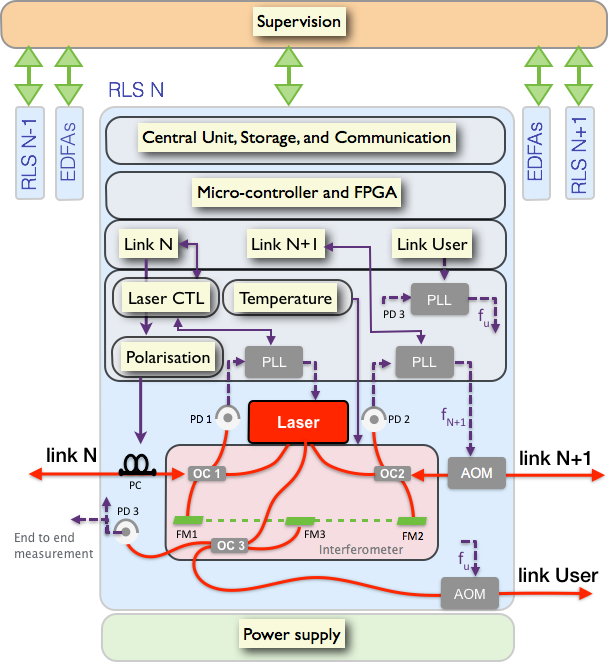}}
  \caption{Functional sketch of the remote controlled repeater laser station. OC: Optical Coupler, FM: Faraday Mirror, AOM: Acousto-Optic Modulator. PC: Polarisation Controller. PD: Photodiode. PLL: phase Lock Loop. CTL: Control. The tracking oscillators used after each photo-detection are not shown in this picture. The pink rectangle represents the temperature controlled box. \label{fig:schema}}
\end{figure}

The concept and principle of operation of RLS are described in details elsewhere\,\cite{lopez_cascaded_2010}. Here, we introduce this optoelectronic device from an engineering system point of view. The functional diagram displayed on Figure\,\ref{fig:schema} shows the main subsystems of a RLS.%

A RLS has three main roles\,\cite{chiodo_cascaded_2015}. First, it regenerates the incoming signal before sending it backward, enabling the compensation of the noise of link $N$. Second, it sends regenerated light to the $N+1$ link and actively stabilizes it. \rev{Third, the repeater laser station allows the dissemination of the compensated signal to a local user. This output can also be used without compensation to measure the end-to-end frequency fluctuations between two RLSs, as it will be described below.}%

A narrow linewidth laser diode (<~5\,kHz) with an output power of approximately 20~mW is at the heart of the opto-electronic device. The laser of station $N$ is combined with the incoming signal of the link to generate a beatnote, optimized using a polarization controller. This beatnote is used to automatically phase-lock the station's laser onto the link signal, so that the phase fluctuations of the link are copied onto the optical phase of the laser. Typically 1 to 2~mW of regenerated laser light travel back in the network fiber. \rev{The optical signal to noise ratio outside the bandwidth of the lock is the one of the laser diode, free of the accumulated spontaneous emission of the EDFAs}. This light will experience on the way back the same delay fluctuations before reaching station $N-1$, where the round trip propagation noise of the link $N$ is detected using a strongly unbalanced interferometer. The propagation noise is corrected by the station $N-1$ using an AOM.

The next span ($N+1$) is fed with 1 to 2~mW of this regenerated laser light. Again the same process occurs, the noise of the next span is compensated after regeneration of the light in RLS $N+1$.%

\subsection{Design and realization}
The repeater laser stations were developed by Muquans in strong interaction with the scientific academic partners and the RENATER network experts, so that both scientific and telecom requirements were taken into account from the design stage.

Muquans developed the physical package and all the optical subsystems, including the control of the laser source. The electronic subsystem is composed of three main modules: the laser control, the interferometer temperature control, and the link and PLL control. The latter was developed by Syrlinks a company specialized in on-board systems for defense and space applications. The RLS includes advanced automated servo loops and an embedded Operating System for local supervision and remote control.

The interferometric setup in the RLS is critical because it houses the optical phase detectors. The stability of the reference arms of the interferometer is essential for a good frequency stability at short and long integration times, because temporal phase fluctuations in these arms will be copied on the link\,\cite{stefani_tackling_2015}. A specific work was conducted on the interferometer, in order to control the optical lengths of the arms. \rev{We implemented an integration procedure to splice the components and control the fiber lengths to a few mm during the assembly of the interferometers. Given a fiber thermal sensitivity of 37~fs/K/m, this ensures that the temperature dependence is greatly reduced. The components were also selected so that the interferometers have reproducible performances from one to the other.} The optical fiber setup is finally housed in a temperature controlled box, which is itself isolated from the environment, \rev{so that results are as much as possible reproductible}. The internal temperature of the interferometer is controlled to better than 0.05$\degree$C for a 1$\degree$C external temperature variation. The temperature of the interferometer is remotely controlled and monitored.%

With this production method, a set of 33 interferometers have already been built and characterized for the REFIMEVE+ project. Average thermal sensitivity is 0.8~fs/K with a standard deviation of 0.7~fs/K, which is comparable to the best values obtained in the prototype stations installed in the Paris-Strasbourg link\,\cite{chiodo_cascaded_2015}, \rev{but with a systematic approach.}%

The electronic processing and control of the signal is an essential subsystem of this technology. A thorough work for ruggedized design was conducted with a careful selection of robust components. The RLS integrates a complex multi-function electronic board with a stand-alone and smart mode of operation. This monolithic board handles all the functionalities required to stabilize a coherent link, such as detecting, amplifying and filtering the signals, tracking the incoming signals, providing the reference and correction signals and automatically adjusting the polarisation.%

The industrial version of the RLS includes all the functionalities and automatic control of the laboratory prototypes, including its cycle-slip detector. In addition, several new functionalities and more control and set points were added.%

The board is equipped with an FPGA and processor, that controls the amplifiers gain, the servo loop parameters, and manages the input/output functions. The board also features monitoring of the temperature, input and output levels of the beatnotes and RF signals. Several check points of the RF processing chain can be monitored remotely and automatically. Fast Fourier Transform of the error signals can be calculated in real time. The bandwidth of the phase lock loops can be measured, saved and exported remotely, allowing for an efficient remote setting.%

The laser device is also remotely operated. \rev{In particular, the RLS features a scanning mode, used to look} for the beatnote with the incoming signal of link $N$.%

An embedded computer manages the serial communications with the subsystems, and enables a back-up communication with a local operator. Events, such as detected cycle-slips, are logged and time-tagged. The computer saves the data on local memory and communicates with servers through an ethernet port and internet protocols.%

An important work was conducted on the thermal management. We optimized the design and selected low power consumption components. This allows to slow down the aging of the electronic boards and reduces perturbations on the optical interferometer. The dimensions and weight were defined taking into account the allocated footprint in the network nodes. RLS is composed of two 19" rack system units, one for the general power supply and the second for the opto-electronic system. The depths are 300~mm and 600~mm, and the heights 2~U and 5~U respectively. The power consumption is 110~W, and the total weight 17~kg.%
 
\subsection{Remote control and supervision\label{SubSec:Supervision}}
\begin{figure}[t]
\centering
\fbox{\includegraphics[width=.95\linewidth]{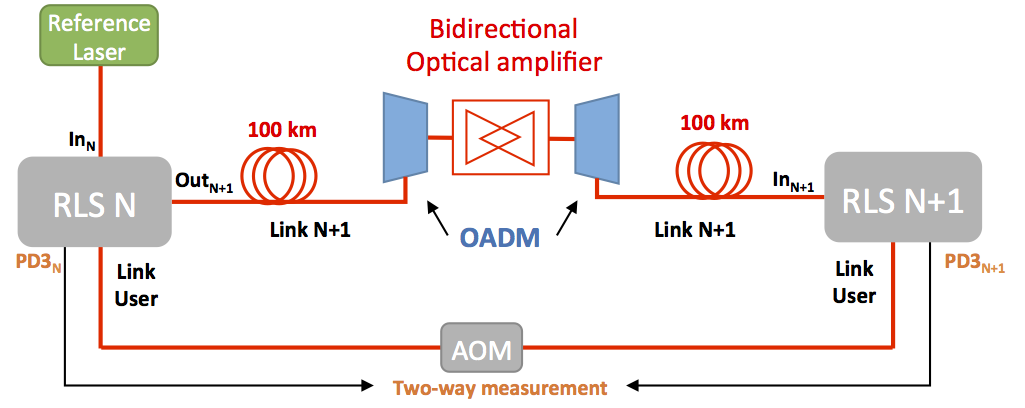}}
\caption{Scheme of the testbench with a simulated link of 200~km, including two bi-directional OADMs and one bi-directional EDFA. For the noise floor measurement, the 200~km link is replaced by a short single mode fiber. }
\label{fig:banc_test}
\end{figure}

Thanks to these features, the automation and remote operation of the RLS have been improved. All the lock-loops and their parameters are digitally controlled and optimized through remote access.%

With this high level of automation, the operation of the link is made in only three top-level steps. First, thermal regulations and the laser are turned on. Then, the RLS checks the incoming signal and locks the laser diode on it. During this step, the polarization is automatically adjusted. The RF gains can also be automatically set if needed. The RLS optimizes all the parameters until the laser is locked. Finally, the $N+1$ and local links are locked when a return signal is detected.%

The RLS is programmed to automatically lock once it is started. If the RLS detects an input signal failure it will run the parameters optimization loop until a satisfying lock is found again. Alarm signals are automatically generated and an operator can remotely control the station in manual mode if necessary.%

Each piece of equipment can be controlled individually. The scientific operators of the network can monitor the status, and check measurement data of all the instruments. They can tune parameters in order to optimize the performance and stability of the metrological network. On top of that a global supervision system was implemented with a dedicated database and several data vizualisation functions. Furthermore, RENATER's Network Operation Center (NOC) has a direct access to supervise, monitor, and control all the equipment. In case of a maintenance operation happening on the network, the NOC can shut down at any time any of the REFIMEVE+ equipment, for the best safety at work and service priority organisation. \rev{Although we have not encountered emergency situations requiring to use this function, we have validated it in our laboratory.} Our work is still in progress, but in a near future the end-users will also be able to connect to the supervision server in order to monitor in real-time the status of the \rev{metrological} service and the validity of the ultra-stable signal delivered to their laboratory.%

\subsection{Operational validation}
\begin{figure}[t]
\centering
\fbox{\includegraphics[width=.95\linewidth]{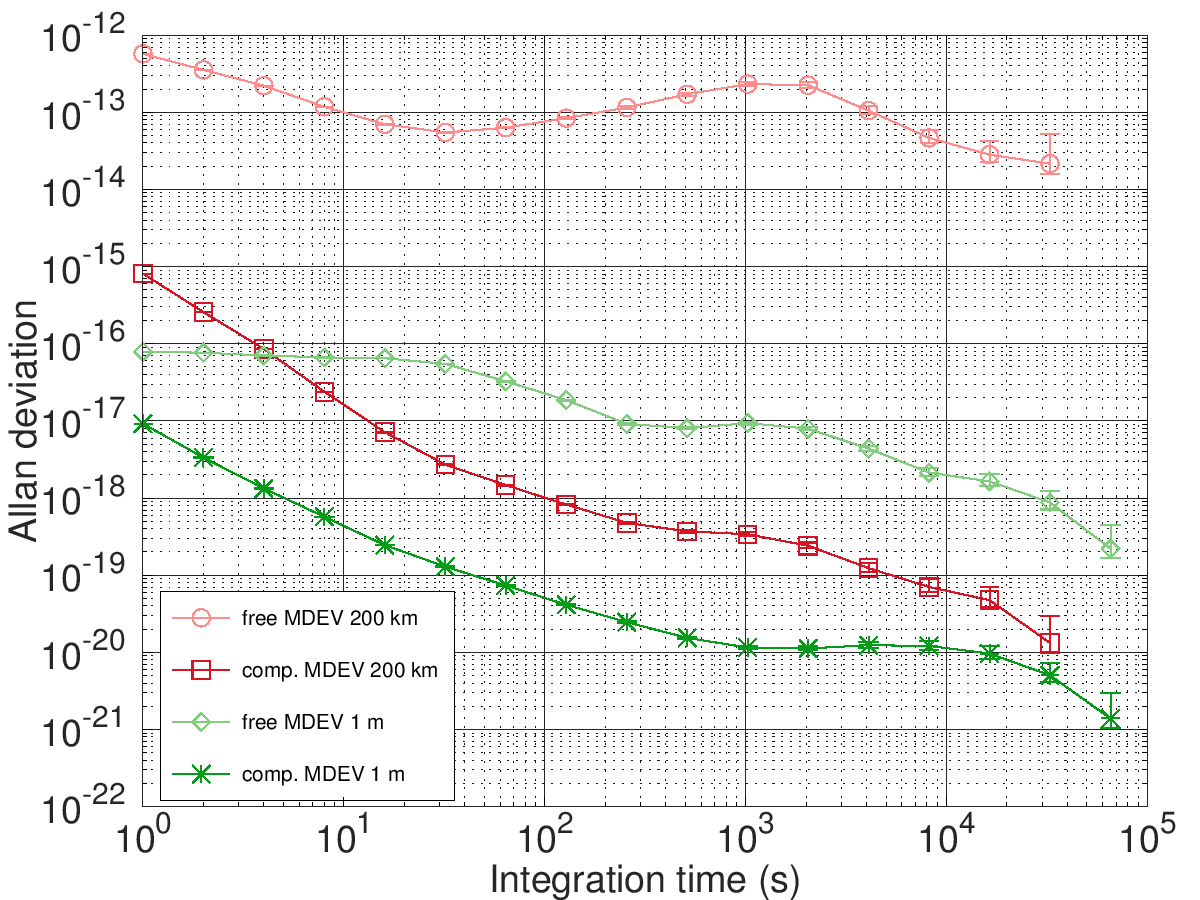}}
\caption{Fractional frequency instability calculated from data recorded with $\Lambda$-counter and 1-s gate time versus averaging time. \rev{Dark red squares: Modified Allan deviation for a compensated 200~km laboratory link. Light red circles: Modified Allan deviation for the free-running 200~km laboratory link. Dark green stars: Modified Allan deviation for a short link to estimate the noise floor of the RLS. Light green diamonds: Modified Allan deviation for the free-running short link.}}%
\label{fig:TW_RS}
\end{figure}

The quality control of each sub-system is carried out before integration by Muquans. Here we focus on the final stage, where the performance of the overall system is assessed.%

Indeed all the RLSs constructed for the REFIMEVE+ project are thoroughly tested in the laboratory before being installed on the RENATER network. At present 22~RLS were manufactured and tested.%

For that purpose, we have built a test bench, sketched out on Figure\,\ref{fig:banc_test}. We use a ultra-narrow laser diode as reference laser (OEWaves Ultra-Narrow Linewidth Laser Module-Gen 3), with frequency noise of about 3~Hz$^{2}$/Hz for Fourier frequency above 1~kHz. The laser frequency is kept close to 194.4 THz and monitored on a regular basis using a calibrated wave-meter.%

We use two repeater laser stations $N$ and $N+1$, \rev{$N$ being the device under test and $N+1$ being used both to send regenerated light backward in the link and to measure its stability.} Station $N$ is locked on the reference laser. As a first step, optimal settings for the laser of this station $N$ are found, then its output port $\textrm{Out}_{N+1}$ is connected to the input port $\textrm{In}_{N+1}$ of the second repeater laser station, either by a short link of 1~m or by a 200~km fiber link made of fiber spools, two OADMs and one bi-directional Keopsys EDFA. On this 200-km fiber link, we expect the overall noise level to be higher than on a real link of the same length because of the phase fluctuations due to the highly correlated temperature changes. Nevertheless, it allows to evaluate the RLS functions over a long fiber link before installation and gives an upper value of the link instability.%

\rev{The ability of the RLS $N$ and $N+1$ to actively compensate the noise of the link is checked by measuring its end-to-end stability. To achieve this, } the user outputs of the two RLSs are connected with a short fiber link that includes an AOM. This setup allows for the measurement of the end-to-end frequency fluctuations \rev{of the link}, using a two-way frequency comparison method as first introduced in \,\cite{calosso_frequency_2014}. The beatnote signal between the laser of the station $N$ and the one transmitted on a short fiber from station $N+1$ and frequency-shifted with the AOM is detected on the $\textrm{PD}3_{N}$ of station $N$. In the same way, the beatnote between the laser signal of station $N+1$ and the one transmitted on a short fiber from station $N$ and frequency-shifted with the AOM is detected on the $\textrm{PD}3_{N+1}$ of station $N+1$. By post-processing these beatnotes, one obtains the phase difference between the user outputs signal of the two stations, which gives the end-to-end phase fluctuations of the link. The beatnote signals between the two laser stations are counted with a dead-time free frequency counter (K+K Messtechnik FXE80) operated in Lambda-mode with a 1~s gate time.%
 
We first measure the noise floor of the RLS under test using a very short link with a few meters of fiber. The resulting Allan deviation is shown on Figure\,\ref{fig:TW_RS}. Frequency stabilities below $2\times 10^{-17}$ at 1~s and a few $10^{-20}$ for $\tau > 10^3$ s are systematically achieved. The small excess of noise above $\tau\simeq$2\,000~s, due to the thermal periodic perturbation of the air conditioning system acting on the station interferometer, is below $2\times10^{-20}$. The long-term instability is as low as $2\times 10^{-21}$ at 1 day averaging time. This instability is well below that of the best optical clocks and shows that the RLS will not limit the link instability at the $10^{-20}$ level. Moreover continuous operation of the stations over several weeks has been demonstrated.%

\begin{figure}[t]
\centering
\fbox{\includegraphics[width=.95\linewidth]{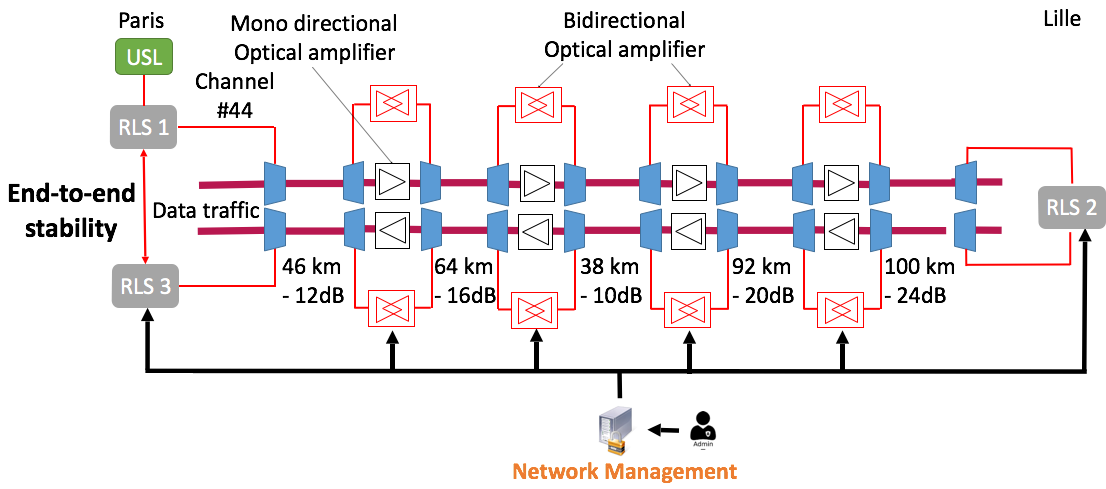}}
\caption{Sketch of the Paris-Lille-Paris link. USL : Ultra-stable laser. Black triangles are the uni-directional EDFA. Double red triangles are the bi-directional EDFA. Blue trapezes are the OADMs. For each span, we provide the span length and the span losses. We remotely supervise and optimize 4 double bi-directional EDFAs and 3 RLSs. Red lines refers to the ultra-stable signal.}%
\label{fig:lien_paris_lille}
\end{figure}

We then replace the short fiber link by the 200~km fiber link on spools. After optimisation of the laser settings, we obtained the results that are shown on Figure\,\ref{fig:TW_RS}. The stability at 1~s is $1\times 10^{-15}$ and decreases to below $10^{-19}$ for $\tau > 10^4$~s, limited by the stability of the reference laser\,\cite{lee_hybrid_2017}. One can again observe a small excess of noise at $\tau\simeq$2\,000~s, due to the thermal periodic perturbation of the air conditioning system acting on the fiber spools bank. Note that this second test stage was only carried out with the first few industrial-grade stations in order to validate all the RLS functions before in-field set-up.%

\section{Deployment of a 680~km optical link : Paris-Lille}
\begin{figure}[t]
\centering
\fbox{\includegraphics[width=.95\linewidth]{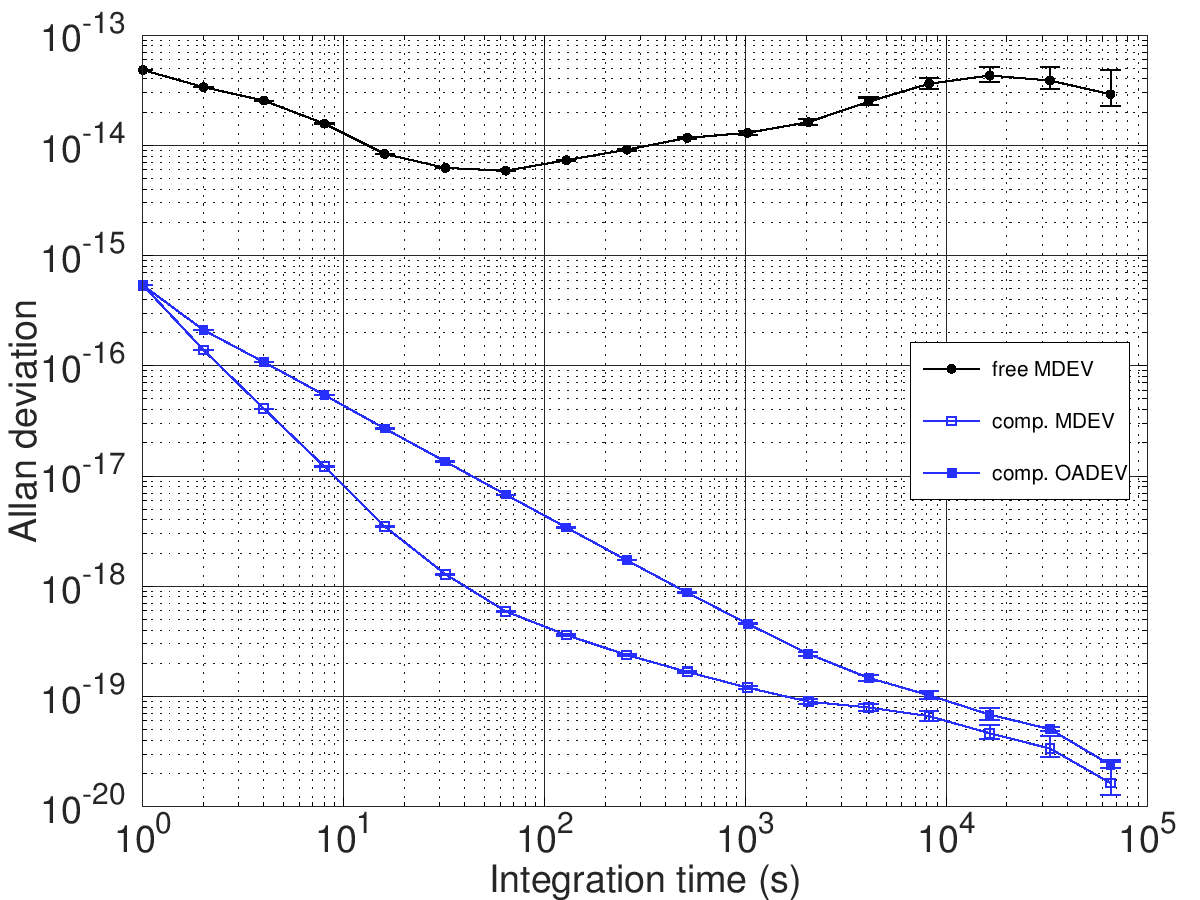}}
\caption{Fractional frequency instability calculated from data recorded with $\Lambda$-counter and 1-s gate time versus averaging time. Black solid circles: Modified Allan deviation for the Paris-Lille free-running link. Blue solid and hollow squares: Overlapping and Modified Allan deviation respectively for the Paris-Lille-Paris compensated link.}
\label{fig:MDEV} 
\end{figure}

The first industrial-grade link of the REFIMEVE+ network was deployed between two cities in France, Paris and Lille, with a geographical distance of about 250~km (Figure \ref{fig:refimeve_carte}). The sketch of the experiment is depicted in Figure\,\ref{fig:lien_paris_lille}. The link is made of two parallel 340~km long fiber spans from the RENATER network, one being for the up link from Paris to Lille and one for the down link back to Paris. In total, the metrological signal travels approximately 680~km. Two RLSs were installed in a RENATER node in Paris at the input and output ends of the link and one RLS in the node at Lille. Four EDFAs were installed in the telecom shelters with 10 OADMs along the link on each span. The optical attenuation on each span is approximately 80~dB. The total gain of the EDFAs is limited to 50~dB on each span of this link in order to prevent self-oscillations in the fiber.%

The 680~km link between Paris and Lille was deployed during the summer of 2017. Installing all the RLSs and EDFAs in the RENATER nodes and shelters took approximately one week. Once all the equipment was set up, remote communication was checked and the link was stabilized, without any malfunction. At that stage, the lasers were locked but many cycle slips degraded the stability and the accuracy of the frequency transfer. The best stability level as shown on Figure\,\ref{fig:MDEV} was reached after further optimizations that lasted approximately 3 weeks. \rev{Because the RLS and EDFAs were fully tested before installation, this period of time was dedicated to adjusting the amplifier and lock loop gains, focusing on the dynamics of the link itself rather than on the equipment.} Overall the installation and adjustment process took under a month. Once the phase lock loops parameters and the gain of the optical amplifiers were optimized, the cascaded link was operated continuously for several weeks. Even when perturbations occur because of thermal fluctuations, acoustic noise or activity on the network, events are time-tagged and the RLSs automatically re-lock in less than one minute. \rev{The interferometer temperature experienced fluctuations with a standard deviation of .065 K during the acquisition of the data.}%

To estimate the stability of the link, we measure the end-to-end signal between the two RLS in Paris, using again a dead-time free frequency counter operated in $\Lambda$-mode with a 1-s gate time. By measuring synchronously the correction signal applied to the AOM of the first station (station 1 in Figure\,\ref{fig:lien_paris_lille}), we can also estimate the free-running noise on the first span of the link. %

Figure\,\ref{fig:MDEV} shows the free-running stability and the Overlapping and Modified Allan deviation of the compensated link during a continuous 5-day measurement. When the link is compensated, we achieve a frequency stability of $5.4\times10^{-16}$ at 1~s integration time, which averages down to $1.7\times10^{-20}$ at 65\,000s. \rev{At short term, it is limited by the propagation delay and at long term by laser and interferometer noise}\,\cite{williams_high-stability_2008,stefani_tackling_2015,lee_hybrid_2017,xu_studying_2018}. This stability is comparable to state-of-the-art ultra-stable optical frequency transfer obtained by research laboratories. 

The accuracy of the frequency transfer is assessed by calculating the mean of the frequency data set, and the statistical uncertainty given by the Overlapping Allan deviation at long integration time\,\cite{benkler_relation_2015}. We took the upper error bar at 65\,000~s for a conservative estimate. One obtains $-3\times10^{-21}\pm(3\times10^{-20})$. Note that the uncertainty is not limited by the inaccuracy of the local RF oscillator\,\cite{chiodo_cascaded_2015}.%

%
\begin{figure}[t]
\centering
\fbox{\includegraphics[width=.95\linewidth]{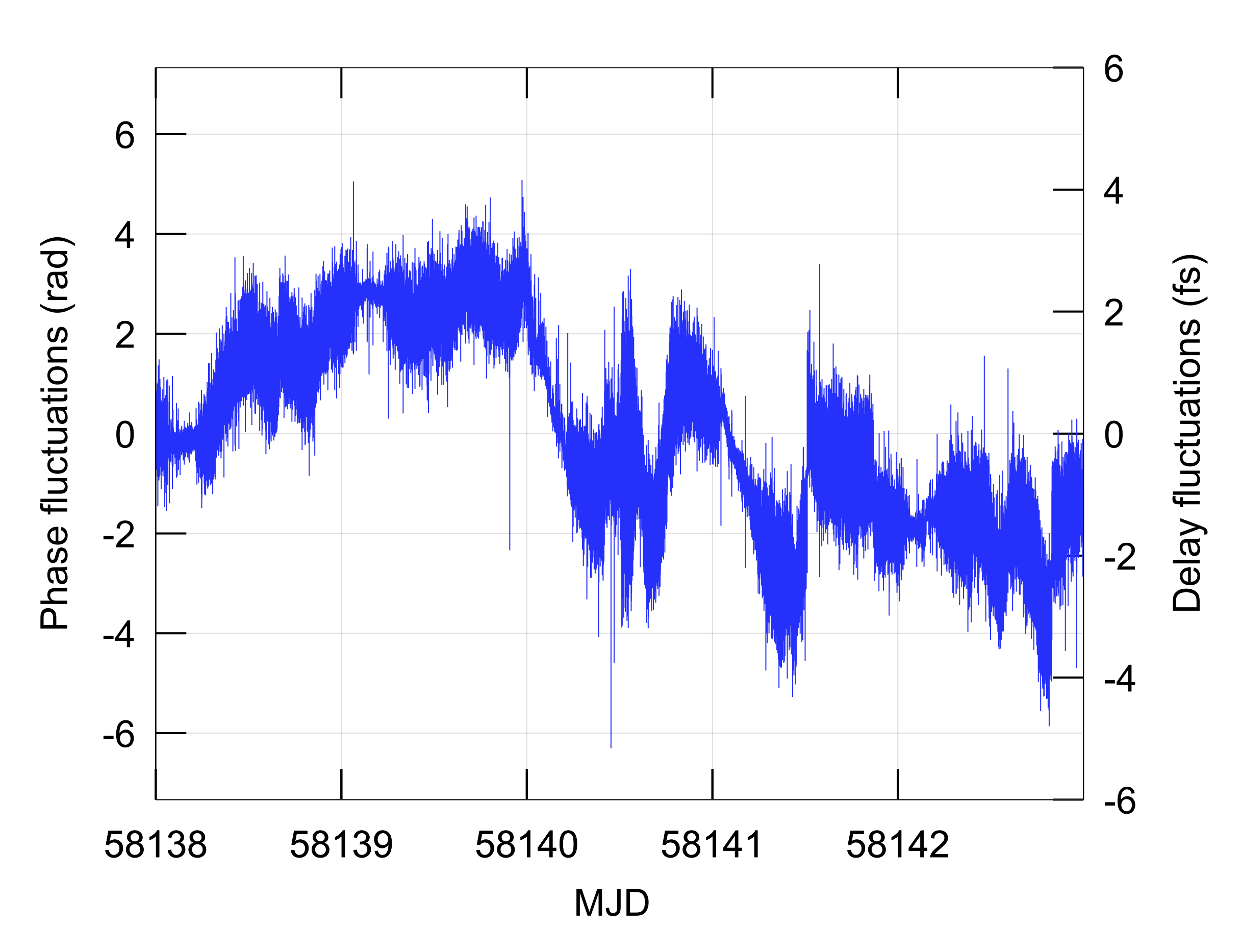}}
\caption{End-to-end phase fluctuations after propagation over a 680~km fiber link.}
\label{fig:phase}
\end{figure}
In order to better appreciate the performances of the link, Figure\,\ref{fig:phase} shows the end-to-end phase fluctuations during this measurement. Over this time period, the RLS tagged 23 cycle slips. Most of them correspond to polarization adjustments. In total 285 data points were removed from the 432\,000~s long data. The integrated phase fluctuations of the end-to end signal are in a range of 10~radians over 5 days. The corresponding time error, expressed in fs, is less than 8~fs. The up-time over this period is as high as 99.93$\%$.

This performance, both in terms of stability and accuracy, meets the requirements of the REFIMEVE+ project.%

\section{Conclusions and perspectives}
We demonstrated the setting-up and exploitation of the first industrial-grade coherent optical fiber link, as part of the REFIMEVE+ network. We show a state-of-the-art relative frequency stability and accuracy at 65\,000~s integration time below the $10^{-19}$ level over a cascaded link of 680 km. Moreover the industrial-grade repeater laser station contribution to the instability is below $2\times10^{-20}$ for integration times higher than 1\,000~s. We also report an uptime of 99.93$\%$ over 5 continuous days, which is very promising regarding the effectiveness of REFIMEVE+ to the end-user.%

With the work presented here, we totally fulfill the requirements of the REFIMEVE+ project. We have demonstrated the basics tools for the upcoming extension of this network. We have also built a test bench to validate the serial production of repeater laser stations.%

Furthermore, we have set-up our production capacity for a full deployment of the network within the next 2 years. The deployment time we can achieve is now comparable to that of a telecommunication link for data traffic, which shows our high level of maturity and concerns about the control of schedule and costs. With this work, the technological readiness level (TRL) of the RLSs was increased from 5-6 to 8. The TRL of the cascaded links with EDFAs and RLSs is now set as high as 8\,\cite{earto_trl_2014}.%

Regarding the hardware developments, the next steps are to reach the same level of maturity for the end-user and eaves-dropping setups. Another major concern is the central hub of the network that is being developed with the concept of super-station\,\cite{cantin_progress_2017}. Finally one crucial challenge relates to the network supervision. Supervision software tools need to be further developed and implemented. We demonstrate here our ability to gather many informations into a complex database. The next challenge is to optimally utilize this database for the most effective network and the best scientific use. Our objective of disseminating frequency standard with high availability is now close to hand.%

\section{Funding Information}
We acknowledge funding support from the Agence Nationale de la Recherche (Equipex REFIMEVE+ ANR-11-EQPX-0039).

\bibliography{Camargo}
 
\end{document}